\documentclass[11pt]{article}
\usepackage[dvips]{graphicx}

\begin{document}

\title{Supersymmetry Breaking, Extra Dimensions and Neutralino Dark Matter}

\author{A.~M. Lionetto\\
Department of Physics and INFN Roma Tor Vergata, \\
Via della Ricerca Scientifica, 1\\
00133 Roma, Italy\\
E-mail: lionetto@roma2.infn.it
}


\maketitle

\begin{abstract}
We show some phenomenological implications for the dark matter problem of a class of models with deflected anomaly mediated supersymmetry breaking in the context of the MSSM. This scenario can be naturally embedded in a brane world model with one compactified extra dimension. It turns out that in these models the neutralino is still the LSP and so a good candidate as cold dark matter. We found that the neutralino is quite a pure bino in almost all the parameter space. Moreover we computed the thermal relic density and we found wide cosmologically allowed regions for the neutralino.
\end{abstract}

\section{Introduction}
Dark matter still remains one of the main unresolved problem in physics. One of the common accepted paradigm is that the solution of the dark matter puzzle involves the existence of an exotic weak interacting massive particle (WIMP). Such a particle has to be found in some extension of the SM of particle physics. It is well known that supersymmetry is an essential ingredient of a consistent theory beyond the SM and the most studied framework is the MSSM, the minimal supersymmetric extension of the SM. In the MSSM the lightest supersymmetric particle (LSP) is usually a neutralino, that is a good candidate as a cold dark matter particle~\cite{Jungman:1995df}. 
The pattern of the soft supersymmetry breaking terms\footnote{for a recent review about soft supersymmetry breaking lagrangian see~\cite{Chung:2003fi}} greatly affects the composition and the strength of the dominant interactions of the neutralino. Hence it is very interesting to study the neutralino phenomenology in different supersymmetry breaking scenarios.
In more recent developments some of these scenarios can also involve
the presence of extra space-time dimensions (see for example~\cite{Randall:1998uk}).  
  
\section{Deflected Anomaly Mediation}
The challenge to find a mechanism able to generate a suitable soft supersymmetry breaking lagrangian without any fine-tuning had a considerable attention during at least the last two decades.
There are essentially three main classes of solutions to this problem. They differ from each other by the primary ``source'' that transmits the supersymmetry breaking to the MSSM fields. 
All these solutions share in common the presence of a hidden sector, that hosts the supersymmetry breaking source, and of a visible sector in which resides the MSSM fields. 
One of the solutions is the well known gravity mediation (see, e.g.~\cite{Nilles:1983ge}) in which the supersymmetry breaking is vehicled by tree-level Planck suppressed couplings. The scenario with minimal coupled $N=1$ supergravity is usually called mSUGRA. The mSUGRA models that satisfies the electroweak symmetry breaking constraints constitute the constrained MSSM (CMSSM)~\cite{Kane:1993td}. Another solution is gauge mediation (see, e.g.~\cite{gaugemediation}) in which the vehicle are the ordinary gauge interactions. 
Finally there is the anomaly-mediated supersymmetry breaking (AMSB) scheme (see, e.g.~\cite{Giudice:1998xp} and~\cite{Randall:1998uk}) in which the supersymmetry breaking is transmitted to the MSSM because of the presence of the superconformal anomaly.
Although very natural, the gravity mediated supersymmetry breaking, in order to be phenomenologically viable, implies a fine-tuned forms for the superpotential and K\"ahler potential~\cite{Randall:1998uk}.
The gauge mediated scheme is not affected by such problem, but on the other hand it very likely predicts the gravitino to be the LSP, which is not the most suitable candidate to constitute cold dark matter.

\begin{figure}[ht]
\begin{center}
\includegraphics[scale=0.4]{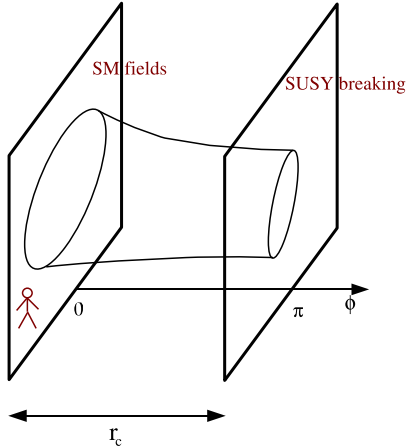}
\caption{Brane configuration in the Randall-Sundrum model. \label{rs-brane}}
\end{center}
\end{figure}

The anomaly mediated models do not have any of the last two undesirable features, since they usually predict the neutralino to be the LSP and allow the possibility to suppress any direct gravitational coupling between the primary supersymmetry breaking source and the MSSM.
This scenario can be naturally embedded in a simple brane world model in a $5$-dimensional theory in which the visible sector fields, i.e. the MSSM fields, are confined on a $3$-brane.
In this kind of models the hidden sector is really a totally sequestered sector, because it resides on a different $3$-brane (see fig.~\ref{rs-brane}).
One of the main results of Randall and Sundrum~\cite{rs1and2} is that in order to have a stable brane configuration the space-time geometry between the two branes has to be a slice of $AdS_5$:
\begin{equation}
ds^2=e^{-2kr_c\phi}\eta_{\mu\nu}dx^\mu dx^\nu+r_c^2 d\phi^2
\end{equation}
where $r_c$ is a ``small'' compactification radius~\cite{rs1and2}, $k$ is a scale of order the Planck scale and $\phi$ parametrize the extra (compact) coordinate. In this way all the massive interaction get suppressed by an exponential factor and only gravity propagates in the bulk. There are no direct contact terms that could lead to phenomenologically dangerous flavor and CP violations effects.

The dynamics in the anomaly mediated sector can be described in terms of a chiral superfield $\varphi=1+F_\varphi\theta^2$, the so-called superconformal compensator superfield~\cite{compensator}.
This is the only field directly coupled to the visible sector. It is worth to remark that the supersymmetry breaking terms are generated at loop level once the auxiliary field $F_\varphi$ acquires a VEV~\cite{Randall:1998uk}.

Anyway the ``bare'' AMSB scheme predicts the uncolored MSSM scalars to be tachyonic and hence it needs for some other mechanism to lift their squared masses to positive values.

A possibility to solve this problem is given by adding another hidden sector~\cite{prepceslio} in the visible brane containing a gauge singlet chiral superfield $X=(A_X,\Psi_X,F_X)$ directly coupled to $N_f$ copies of messenger chiral superfields $\Phi_i$, with $i=1,\ldots,N_f$. These fields are assumed to transform under the fundamental and anti-fundamental of the standard MSSM gauge groups as in the usual gauge mediated scenario. 
Due to the superconformal anomaly, the supersymmetry breaking, originated in the hidden sector, is communicated both to the standard models fields and to the gauge mediated hidden sector fields.
We assumed that the supersymmetry breaking effects are transmitted to the gauge mediated sector at tree level and that the superpotential of the gauge singlet $W\left(X\right)$ must contain only terms with couplings of positive or vanishing mass dimension:
\begin{equation}
W\left(X\right)=a_1 X^3+a_2 \left<F_\varphi\right>X^2+a_3 \left<F_\varphi\right>^2 X+a_4 \left<F_\varphi\right>^3,
\label{wx}
\end{equation}
where $a_1, a_2, a_3, a_4$ are dimensionless real parameters.

The compensator auxiliary field VEV $\left<F_\varphi\right>$ (that is coupled to every field in the visible brane) induces nonzero VEVs for both the scalar and auxiliary parts of the $X$ superfield, $\left<A_X\right>$ and $\left<F_X\right>$.
In this way the messenger superfields $\Phi_i$ and ${\bar{\Phi}}_i$ acquire masses~\cite{gaugemediation} of the order of~$\left<A_X\right>$ and mass splittings of the order of~$\sqrt{\left<F_X\right>}$.
Furthermore the presence of the intermediate threshold given by the nonzero VEV of the $X$ superfield will \emph{deflect}\footnote{for other explicit realization of this class of models see~\cite{Rattazzi:1999qg} and~\cite{Okada:2002mv}.} the renormalization group (RG) trajectories of the soft supersymmetry breaking terms off the AMSB trajectory.
This deflection is able to eliminate the negative squared masses for the uncolored MSSM scalars, that are present in the minimal realization of the AMSB scenario.

\section{Neutralino and Deflected Anomaly Mediation}
The phenomenology of the deflected scenario is quite distinctive due to presence of an intermediate energy scale $\left<A_X\right>=\xi F_\varphi$, where $\xi$ is a new dimensionless parameter, that sets the typical mass of the messenger superfields.
It is possible to derive the usual soft supersymmetry breaking terms~\cite{prepceslio} which depend by the following parameters
\begin{equation}
\left<X\right>=m(1+\theta^2 f/m),\quad \left<\varphi\right>=1+\theta^2 F_\varphi
\end{equation}
where we have defined
\begin{eqnarray}
m & = & \left<A_X\right>=\xi F_\varphi\\
f & = & \left<F_X\right>=d\xi F_\varphi^2
\label{mf}
\end{eqnarray}
The parameter $d$ indicates how much the RG anomaly mediated trajectory is deflected 
\begin{equation}
\frac{f}{m}=d F_\varphi
\end{equation}
It is possible to show that a superpotential of the type of eq.~\ref{wx} implies $d>0$. Thus this scenario is usually termed as positively deflected anomaly mediated~\cite{Okada:2002mv}.
The boundary conditions for the soft terms RG equations are given at the renormalization scale $\mu=m=\xi F_\varphi$. The weak scale $M_Z$ predictions are then obtained by using the usual MSSM renormalization group equations at two loop level~\cite{Martin:1993zk}.
To perform the running we used the ISASUGRA RGE package, which is contained in the ISAJET 7.64 package~\cite{Baer:1999sp}.
For the computations of all the relevant quantities at the weak scale we use the DarkSUSY code~\cite{Gondolo:2002tz}. 
As we have already seen the soft term expressions are entirely determined by the two mass parameters $F_\varphi$ and $\left(f/m\right)$ and by the dimensionless number $N_f$.
It is then possible to study the phenomenological properties of the
deflected model through contour plots in the
$\left(f/m,F_\varphi\right)$ plane fixing $\xi$, which determines the scale at which the boundary conditions are given ($m=\xi F_\varphi$), the ratio between of the two Higgs VEVs $\tan \beta$, the sign of the Higgs $\mu$ term and the number of messenger flavors $N_f$.
It turns out that the neutralino is the LSP in almost all the parameter space.
This is a consequence of the fact that the fermionic component $\psi_X$ of the hidden sector scalar superfield $X$ has a mass of the order $F_\varphi$.
Moreover the gravitino mass is $m_{3/2}\sim F_\varphi$~\cite{Randall:1998uk} and all the soft breaking masses are suppressed by the square of the gauge couplings.
Thus neither the fermionic component $\psi_X$ of the gauge singlet
superfield nor the gravitino are the lightest supersymmetric
particle. The neutralino remains as the main LSP candidate as in the mAMSB scheme. 
\begin{figure}[ht]
\includegraphics[scale=0.55]{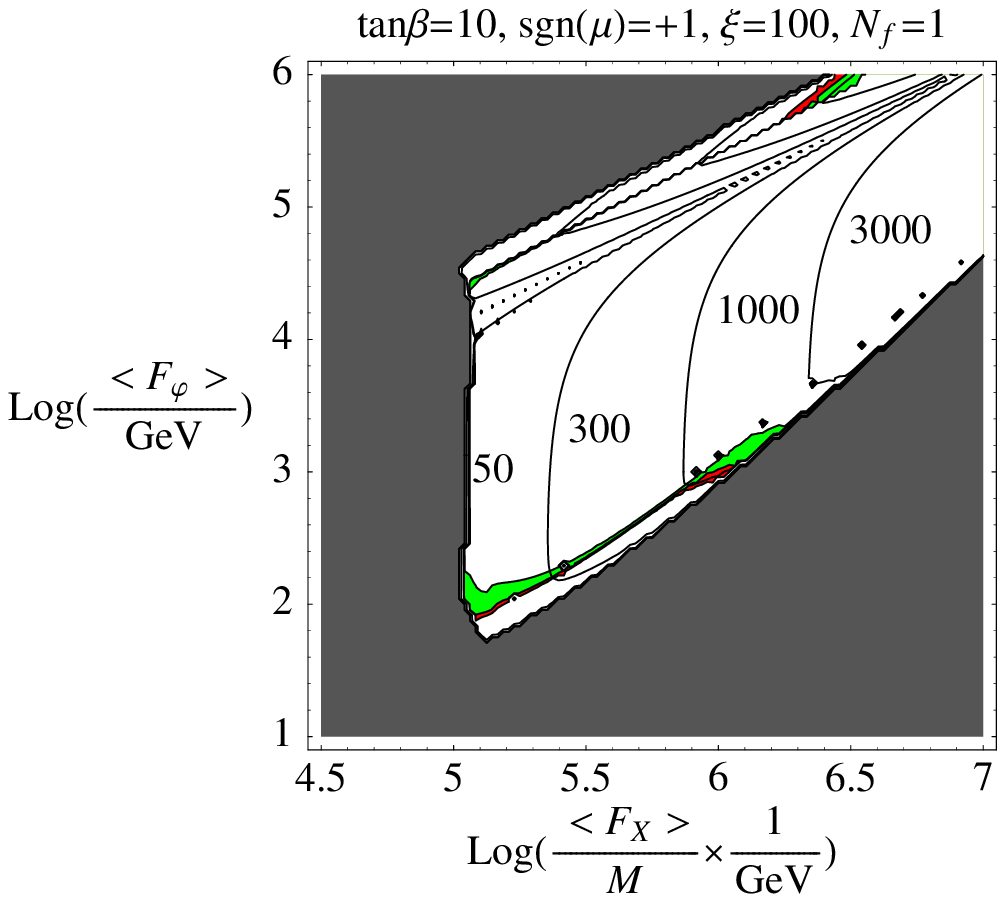}
\includegraphics[scale=0.55]{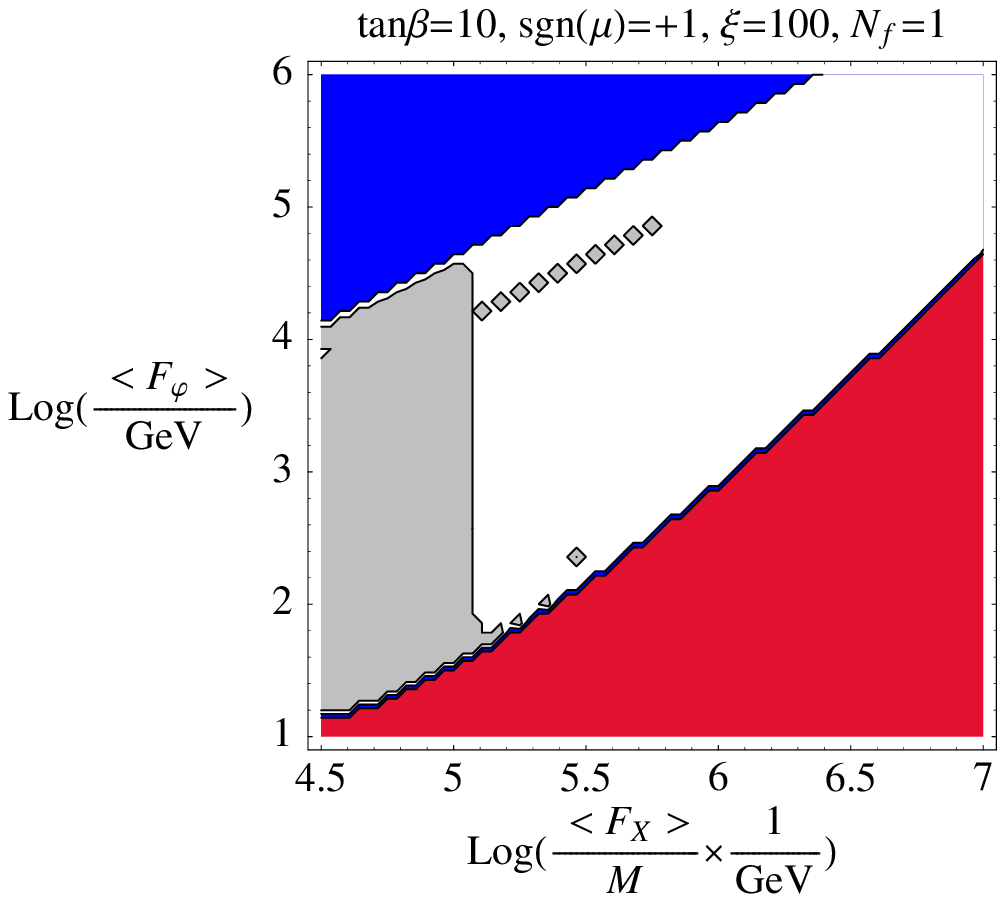}
\caption{Left panel: neutralino mass $m_\chi$ (expressed in GeV) contour plot. The cosmologically allowed regions are in red (WMAP range: $0.09<\Omega h^2<0.13$) and in green ($0.13<\Omega h^2<0.3$) Right panel: exclusion zone (see the text for detailed explanation).\label{mchi_mixed_tb10}}
\end{figure}

In the left panel of fig.~\ref{mchi_mixed_tb10} we show the neutralino isomass $m_\chi$ (expressed in GeV) contour plot in the plane ($f/m$,$F_\varphi$) together with the cosmologically allowed zone: the red region is the one allowed by the WMAP data~\cite{Spergel:2003cb} $0.09<\Omega h^2<0.13$ while in the green one $0.13<\Omega h^2<0.3$. The dark gray region is the excluded one due to different reasons. 
In the right panel of fig.~\ref{mchi_mixed_tb10} we show the details
of the excluded region: the blue region is excluded due to the
presence of tachyons, the light gray one is excluded by LEP bounds
violations while the red one is excluded because there is no electroweak symmetry breaking.

In the deflected anomaly scenario the neutralino is a very pure bino in almost all the parameter space. This is the main difference with the mAMSB scenario in which the neutralino is in general wino-like. The range of possible masses for the neutralino is wider than in the well studied mSUGRA case.
Unlike the latter case it is possible to have a light neutralino of few GeV because the LEP constraints are no more valid here~\cite{lightneutralino} due to the non universality of the gaugino masses at the boundary scale (see eq.~\ref{mgauginiexpl} in appendix~\ref{deflected_gaugino}). 
Moreover there are regions in which the neutralino has a mass $m_\chi\sim 1$ TeV and also the right cosmological abundance.

\begin{figure}[ht]
\includegraphics[scale=0.55]{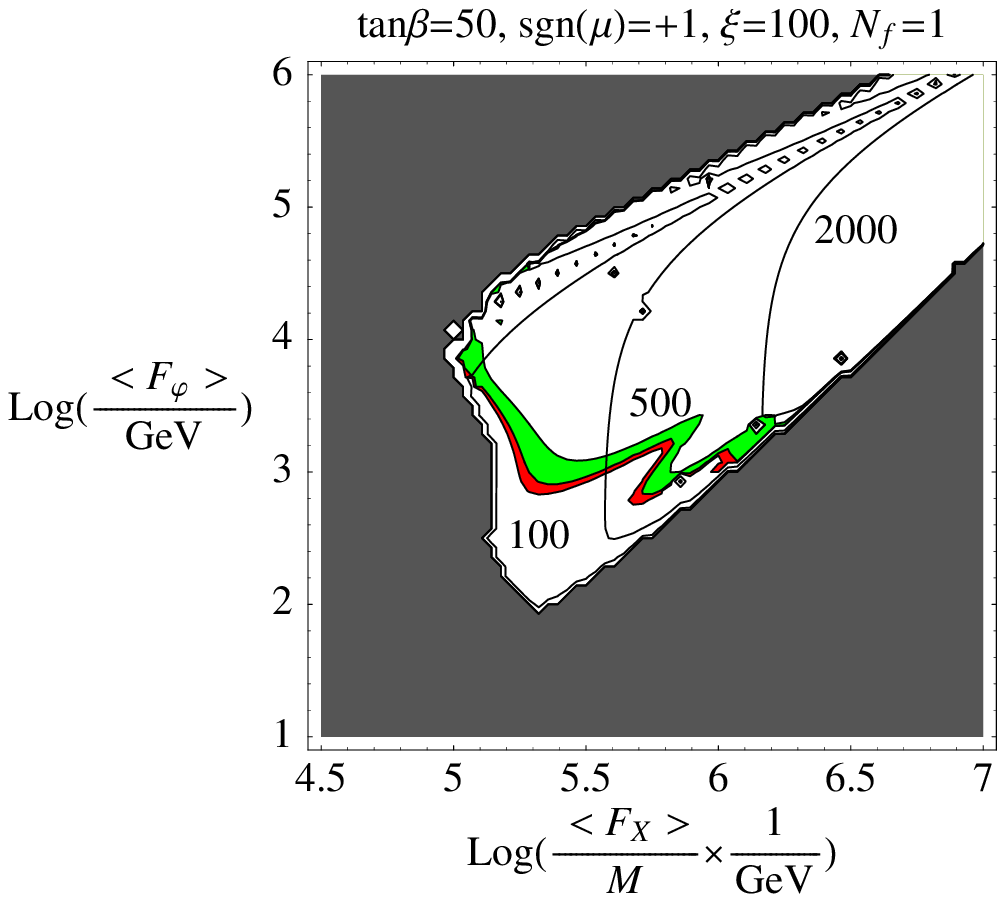}
\includegraphics[scale=0.55]{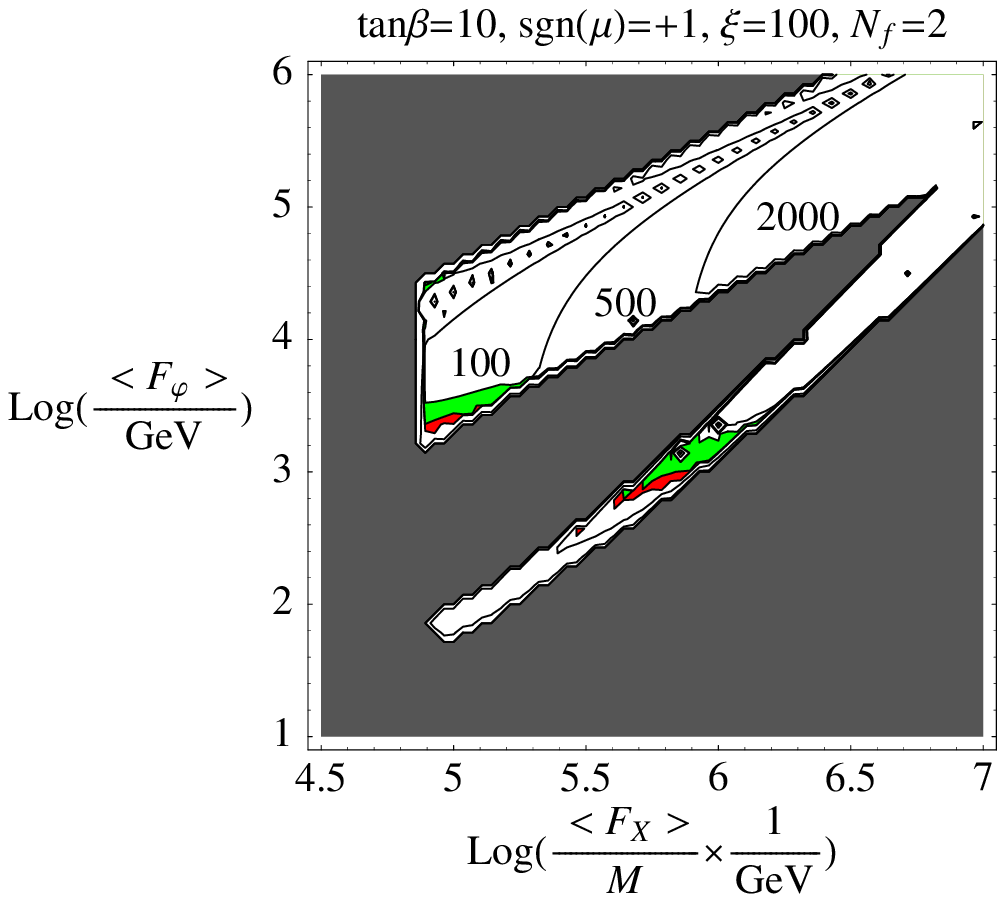}
\caption{Left panel: neutralino mass $m_\chi$ (expressed in GeV) contour plot for $\tan\beta=50$ and $N_f=1$ together with the cosmologically allowed regions (see fig.\ref{mchi_mixed_tb10}). Right panel: the same for $\tan\beta=10$ and $N_f=2$.\label{mchi_mixed_tb50_and_nf2}}
\end{figure}

In fig.~\ref{mchi_mixed_tb50_and_nf2} there are other two neutralino isomass contour plots for a different choice of the parameters $\tan\beta$ and $N_f$. In particular in the left panel $\tan\beta=50$ and $N_f=1$ while in the right panel $\tan\beta=10$ and $N_f=2$. 
It is worth noting that for $N_f>1$ the parameter space is more constrained with respect to the case $N_f=1$, i.e. the allowed parameter space is smaller.

\section{Conclusions}
We presented a peculiar class of models with (positive) deflected anomaly mediated supersymmetry breaking that can be naturally embedded in a brane-world five-dimensional scenario. These kind of models are able to solve the AMSB tachyon problem. 
Moreover they constitute an interesting alternative scenario in the context of the dark matter problem. We showed that in a deflected anomaly scenario the neutralino is still the LSP and tends in general to be a very pure bino. 
We found that the neutralino has the right relic abundance to constitute cold dark matter in a wide portion of the parameter space, if compared to other supersymmetry breaking scenarios in which the neutralino is the LSP, as for example the gravity-mediated ones.  
 
\section*{Acknowledgments}
The author would like to thank Nicolao Fornengo and Stefano Scopel for very useful discussions.

\appendix

\section{Explicit Expressions for the Gaugino Terms\label{deflected_gaugino}}
We show as an example the prediction for the gaugino mass term $M_i$ ($i=1,2,3$) for a (positively) deflected anomaly mediated scenario. The terms are computed at the renormalization scale $\mu=m$, the typical scale of the messenger fields:
\begin{equation}
\left\{
\begin{array}{lc}
\left.M_{1}\right|_{\mu=m}=-\frac{g_1^2(m)}{(4\pi)^2}\left[N_f\frac{f}{m}+(-\frac{33}{5}-N_f)F_\varphi\right] \\ \\
\left.M_{2}\right|_{\mu=m}=-\frac{g_2^2(m)}{(4\pi)^2}\left[N_f\frac{f}{m}+(-1-N_f)F_\varphi\right] \\ \\
\left.M_{3}\right|_{\mu=m}=-\frac{g_3^2(m)}{(4\pi)^2}\left[N_f\frac{f}{m}+(3-N_f)F_\varphi\right]
\end{array}
\right.
\label{mgauginiexpl}
\end{equation}
where the $g_i$ ($i=1,2,3$) are the corresponding gauge couplings. 
It is interesting to note that these terms are actually the sum of an anomaly mediated and a gauge mediated contribution and that they are not universal. This fact holds only at the messenger scale.

\end{document}